# FLASH: Ultrafast beam quality characterization via spatial-to-temporal mapping

JUNJIE QIU, YUXUAN XIONG, XUCHEN HUA, HAO WU* AND MING TANG*

*Wuhan National Laboratory for Optoelectronics (WNLO), School of Optical and Electronic Information, Huazhong University of Science and Technology, Wuhan 430074, China*
** wuhaoboom@hust.edu.cn and tangming@hust.edu.cn*

**Abstract:** Accurate and real-time monitoring of spatial beam quality has emerged as the absolute prerequisite for intelligent optical field regulation and advanced laser applications. However, modern high-power and multimode optical systems exhibit highly complex, nonlinear, and transient behaviors. In these systems, the spatial beam profile undergoes dramatic reorganizations within extremely short timeframes. Phenomena such as spatio-temporal mode-locking, transient beam self-cleaning, and plasma-induced aberrations demand nanosecond-level dynamic characterization. Yet, capturing these ultrafast dynamics is fundamentally bottlenecked by the kilohertz frame rates of conventional two-dimensional image sensors. To break this dimensional and temporal barrier, we propose an ultrafast non-imaging beam quality monitoring technique, termed Fiber-based Laser Assessment via Spatial-to-temporal High-speed-mapping (FLASH). By utilizing a multimode fiber to encode spatial beam variations into high-dimensional speckle fingerprints and a multicore fiber delay line array to serialize these features, we transform two-dimensional spatial information into high-speed one-dimensional temporal pulse sequences. Empowered by a deep learning model to decipher the serialized signals, the FLASH system achieves an unprecedented 100 MHz measurement rate with a minimal mean relative error of 0.32%. Realizing a five-order-of-magnitude speed improvement over standard camera-based methods, this spatial-to-temporal mapping paradigm provides a transformative spatial oscilloscope. It unlocks new possibilities for real-time intelligent adaptive control and the exploration of complex multimode nonlinear physics.

## 1. Introduction

The continuous evolution of laser technology has driven a transition from the fundamental pursuit of output power to the sophisticated realm of intelligent optical field regulation. From customized spatial beam profiles for defect-free welding in extreme precision manufacturing [1-3] to multi-channel data transmission in spatial-multiplexed optical communications [4-6] and long-distance high-energy directed systems [7-9], the precise manipulation of spatial beam properties has become the cornerstone of various advanced applications. To fully harness the potential of these applications, realizing closed-loop adaptive optical control is imperative. This goal fundamentally depends on the accurate and real-time monitoring of beam quality. However, implementing such intelligent monitoring and control is severely challenged by the intrinsic characteristics of modern high-power and multimode optical systems.

In these systems, the optical field is rarely static, exhibiting highly complex, nonlinear, and transient behaviors. In terms of exploring multimode nonlinear physics, spatial profile distortions reflect the complex interplay between transverse modes and longitudinal pulses. Here, nanosecond-level dynamic characterization of beam quality serves as the direct basis for unveiling multimode spatio-temporal dynamics such as spatio-temporal mode-locking [10-13] and transient beam self-cleaning [14-17]. Regarding the support for intelligent adaptive control, whether suppressing Transverse Mode Instability (TMI) in large-mode-area (LMA) fiber lasers to maintain the stable operation of ultra-high-power systems [18-20] or compensating for plasma-induced transient aberrations during laser processing [21-22}, there is an urgent reliance on spatial beam quality feedback at extremely short timescales. In extending the limits of free-space transmission, real-time and ultrafast beam quality monitoring is the core technical means to combat high-frequency atmospheric vortex perturbations and overcome turbulence-induced distortions [23,24]. Because these intricate nonlinear processes cause the spatial beam profile to undergo dramatic reorganizations within an extremely short timeframe, the continuous monitoring of spatial beam quality at an ultrafast rate is an absolute prerequisite for unveiling these hidden high-speed physical mechanisms and achieving intelligent optical field regulation.

Currently, the beam quality factor, universally known as the $M^2$ factor, serves as the standard metric for quantifying such dynamic laser propagation characteristics [25-28]. Defined by the ISO 11146 standard, the $M^2$ factor represents the ratio of the beam parameter product (the product of the waist radius and the far-field divergence half-angle) of an actual beam to that of an ideal fundamental Gaussian beam ($TEM_{00}$) at the same wavelength [29]. An $M^2$ value of unity indicates an ideal diffraction-limited beam, while higher values signify degradation in beam quality. While accurate and real-time monitoring of the $M^2$ factor is critical for the aforementioned high-speed applications, traditional measurement techniques fall short. According to the ISO standard, accurate determination requires measuring the beam width at multiple positions along the propagation axis. Conventionally, the variable-distance method employs a motorized translation stage to move a charge-coupled device (CCD) or a knife-edge detector to scan the beam caustic. While this multi-point fitting approach yields high precision, its inherent slowness (often taking minutes) renders it unsuitable for monitoring dynamic lasers. To circumvent longitudinal scanning, variable-focus techniques utilizing tunable liquid lenses or spatial light modulators (SLMs) have been developed to scan the focal spot at a fixed plane [30-32]. However, these methods remain constrained by the limited response speed of active optical elements and often suffer from reduced resolution and sensitivity to thermal fluctuations.

To eliminate mechanical motion and accelerate acquisition, computational reconstruction methods based on single-plane measurements have been extensively explored. One category involves virtual propagation. In this method, the complex optical field (amplitude and phase) is first retrieved typically using techniques like lateral shearing interferometry. Subsequently, the field is numerically propagated using diffraction theory to reconstruct the entire beam caustic [33-36]. Additionally, modal decomposition (MD) has gained prominence. This method decomposes the beam into a set of orthogonal eigenmodes to analytically derive beam quality [37-39]. Numerical algorithms such as Stochastic Parallel Gradient Descent (SPGD) are typically employed to resolve modal weights and phases from intensity patterns. Although these numerical approaches significantly reduce system complexity compared to mechanical scanning, they heavily rely on complex phase retrieval or iterative calculations. This introduces inherent processing latency, restricting their applicability to static or slowly varying beams rather than ultrafast dynamic events.

Recently, the integration of deep learning (DL) has marked a paradigm shift in beam characterization. By training Convolutional Neural Networks (CNNs) to map near-field intensity images directly to modal parameters, researchers have achieved accurate modal decomposition and $M^2$ estimation at speeds significantly surpassing conventional numerical methods [40-43]. For instance, shallow neural networks have been demonstrated to estimate the $M^2$ factor in few-mode fibers with high efficiency [44]. Despite these algorithmic breakthroughs enabling millisecond-scale inference, a fundamental physical bottleneck persists. Virtually all existing DL-based methods rely on 2D image sensors (CCD or CMOS) for data acquisition. The measurement rate is strictly capped by the camera frame rate (typically < 1 kHz). This gives rise to a profound diagnostic challenge. While temporal laser dynamics can be readily resolved at gigahertz bandwidths using modern oscilloscopes, the observation of spatial dynamics is fundamentally bottlenecked by these sluggish frame rates. This tremendous mismatch between temporal and spatial measurement capabilities leaves critical ultrafast spatial phenomena completely unresolved.

To enable high-speed measurements beyond the frame rate limitations of 2D image sensors, transforming the spatial beam profile into a serially sampleable format is essential. In this context, the utilization of Multimode Fibers (MMFs) as diffractive encoding elements offers a distinct advantage. Within an MMF, the incident light excites a multitude of guided modes that interfere during propagation to form a complex speckle pattern at the distal end. This speckle distribution serves as a high-dimensional and deterministic fingerprint that is extremely sensitive to the spatial phase and amplitude variations of the input beam. It effectively encodes the intrinsic beam quality information into the intensity domain. Unlike conventional free-space propagation, this all-fiber modal mixing mechanism compresses the rich two-dimensional spatial information into a specific format. This format can then be readily serialized by our subsequent multicore fiber array. When coupled with deep learning algorithms that excel at deciphering complex non-linear mappings, MMFs provide a robust physical foundation for realizing ultrafast and non-imaging beam characterization.

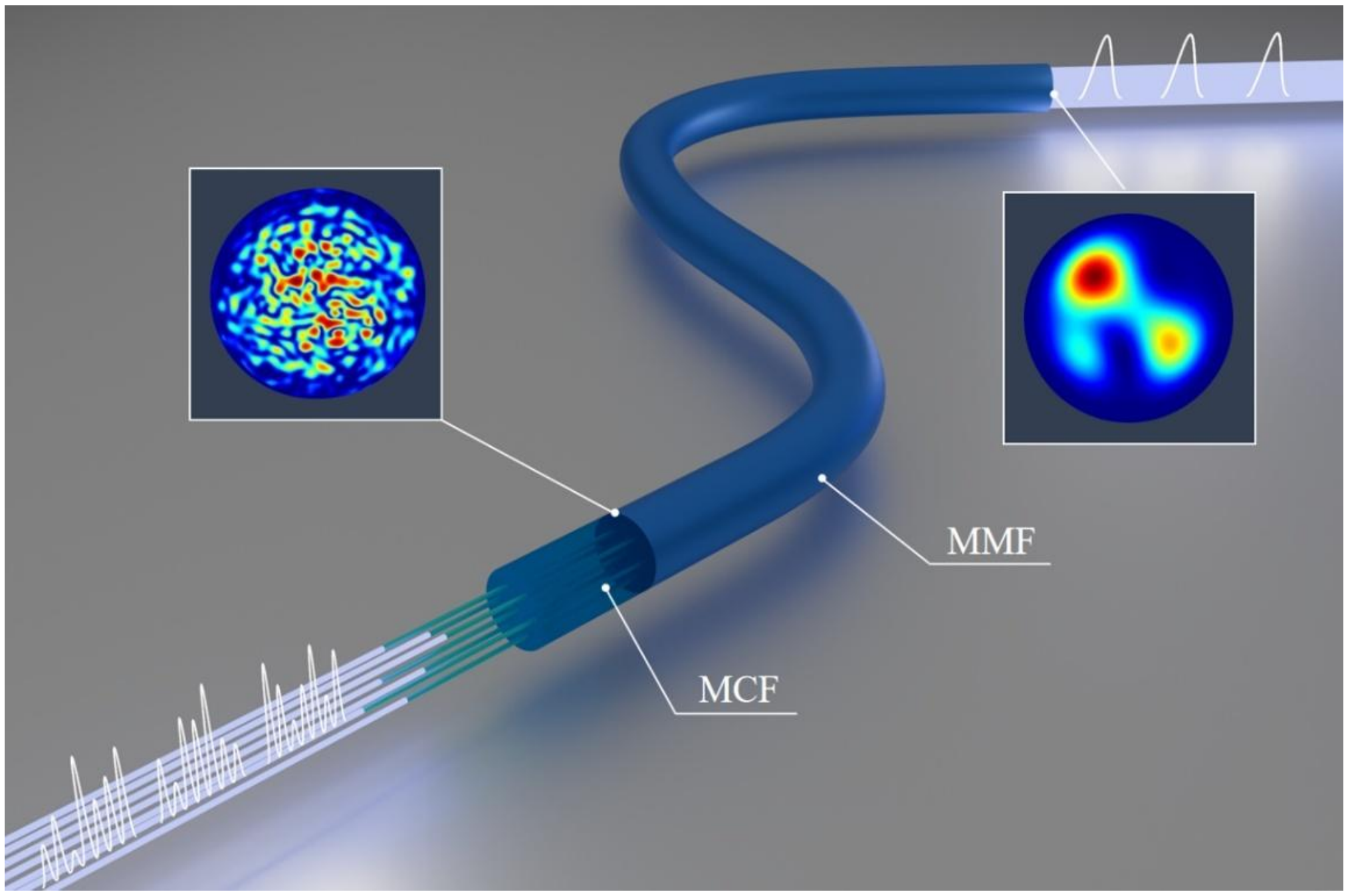

Fig. 1. Principle of the spatial-to-temporal mapping technique. The spatial information of the input light pulse is encoded into a complex speckle pattern via intermodal interference within the Multimode Fiber (MMF). This 2D spatial distribution is subsequently sampled by a Multicore Fiber (MCF) and serialized into a 1D high-speed temporal pulse sequence through fiber delay lines of varying lengths, preventing temporal overlap.

Capitalizing on this diffractive encoding capability, we propose an ultrafast $M^2$ factor measurement technique [45], termed Fiber-based Laser Assessment via Spatial-to-temporal High-speed-mapping (FLASH). As illustrated in Fig. 1, in this scheme, the spatial information of the laser beam is first transformed into unique speckles via intermodal interference in an MMF. Crucially, to bypass the frame rate limitations of conventional 2D sensors, this spatial distribution is subsequently serialized into a high-speed temporal pulse sequence using a multicore fiber (MCF) delay line. This signal is detected by a single photodetector, allowing for data acquisition speeds far beyond the reach of cameras. Integrated with a shallow neural network for efficient regression, the FLASH system achieves a measurement rate of 100 MHz with a mean relative error of 0.32%. Realizing a five-order-of-magnitude speed improvement over standard camera-based methods, this spatial-to-temporal mapping paradigm successfully bridges the long-standing gap between slow spatial imaging and ultrafast physics. It provides a transformative spatial oscilloscope for capturing nanosecond-level transient optical phenomena and offers a robust solution for the high-precision real-time characterization of ultrafast laser dynamics.

## 2. Materials and methods

### *2.1. Experimental setup*

The schematic diagram of the proposed FLASH system is illustrated in Fig. 2. The experimental setup comprises three primary functional modules: the programmable beam generation module, the spatial-to-temporal conversion module, and the high-speed signal acquisition module.

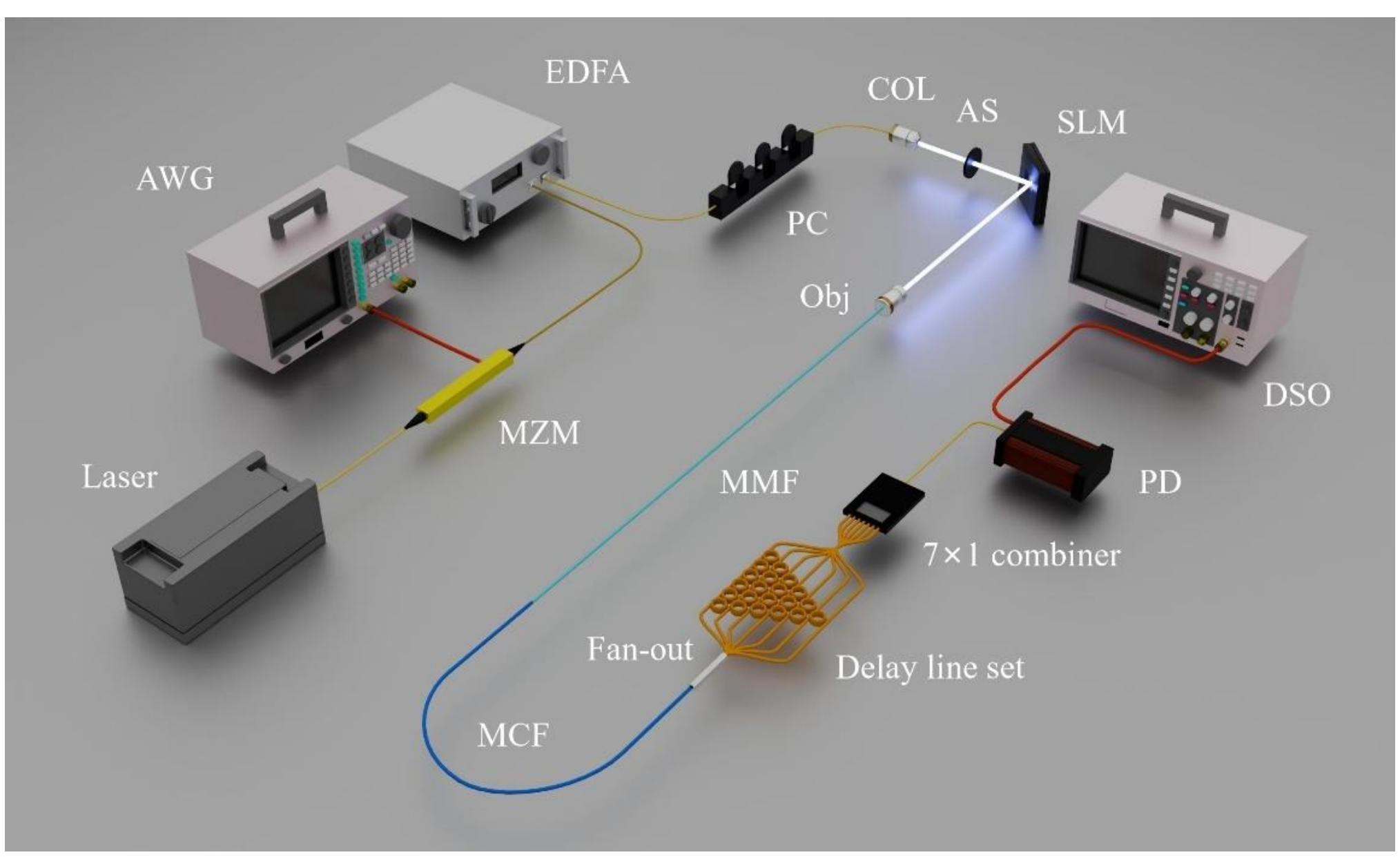

Fig. 2. Schematic diagram of the experimental setup. The system generates programmable beam profiles and performs ultrafast detection. Key components include: AWG, arbitrary waveform generator; MZM, Mach-Zehnder modulator; EDFA, erbium-doped fiber amplifier; PC, polarization controller; COL, collimator; AS, aperture stop; SLM, spatial light modulator; PD, photodetector; DSO, digital storage oscilloscope.

In the beam generation module, a narrow linewidth laser operating at 1550 nm serves as the light source. The continuous-wave output is modulated into high-speed optical pulses via a Mach-Zehnder modulator (MZM) driven by an arbitrary waveform generator (AWG). An erbium-doped fiber amplifier (EDFA) is employed to amplify the optical pulses to compensate for transmission losses. Regulated by a collimator (COL) and an aperture stop (AS), a collimated spatial light pulse of fixed size is projected onto a phase-only spatial light modulator (SLM). By loading specific phase maps corresponding to the near-field images, we can generate diverse beam profiles with controllable beam quality to emulate various laser states. The construction of the corresponding ground-truth dataset are detailed in Section 2.2. A polarization controller (PC) is placed before the SLM to ensure the polarization state of the incident light matches the sensitivity of the liquid crystal molecules.

The modulated optical pulses are subsequently coupled into the spatial-to-temporal conversion module, which constitutes the core of our proposed method. The beam is focused by a 20× objective lens into a large-core step-index MMF (core/cladding diameter: 200/220 $\mu$m). Inside the MMF, the incident light excites a multitude of guided modes, which propagate and interfere to form a complex speckle at the distal end. This speckle acts as a high-dimensional fingerprint of the input beam quality. To serialize this 2D spatial information, the output end of the MMF is center-aligned and spliced to a 7-core MCF (cladding diameter: 150 $\mu$m, core diameter: 8 $\mu$m, core-to-core spacing: 42 $\mu$m). The MCF spatially samples the speckle field at seven discrete positions. The MCF is then connected to a fan-out device, which separates the seven cores into individual single-mode fibers. To map the spatially sampled signals into the time domain, a fiber delay line array is constructed to ensure that optical pulses from different cores arrive at the detector sequentially without temporal overlap.

Finally, in the signal acquisition module, the seven time-delayed optical signals are merged into a single path using a 7 × 1 fiber combiner. The combined signal is detected by a high-speed single-pixel photodetector (PD) with a bandwidth of 20 GHz. Consequently, the spatial intensity distribution sampled by the MCF is converted into a high-speed temporal pulse sequence. An additional EDFA is utilized to compensate for optical transmission losses. The electrical output from the PD is acquired by a Digital Storage Oscilloscope (DSO) with a sampling rate of 20 GSa/s for subsequent data processing.

### *2.2. Generation of near-field beam profiles and spatial dataset construction*

To drive the programmable beam generation module described in Section 2.1 and to train the neural network for high-precision $M^2$ prediction, a large-scale dataset containing diverse beam profiles and their corresponding ground-truth $M^2$ labels is essential. Since acquiring massive labeled data from experimental measurements is time-consuming and prone to calibration errors, we employed a numerical simulation approach based on the theory of modal decomposition and beam propagation.

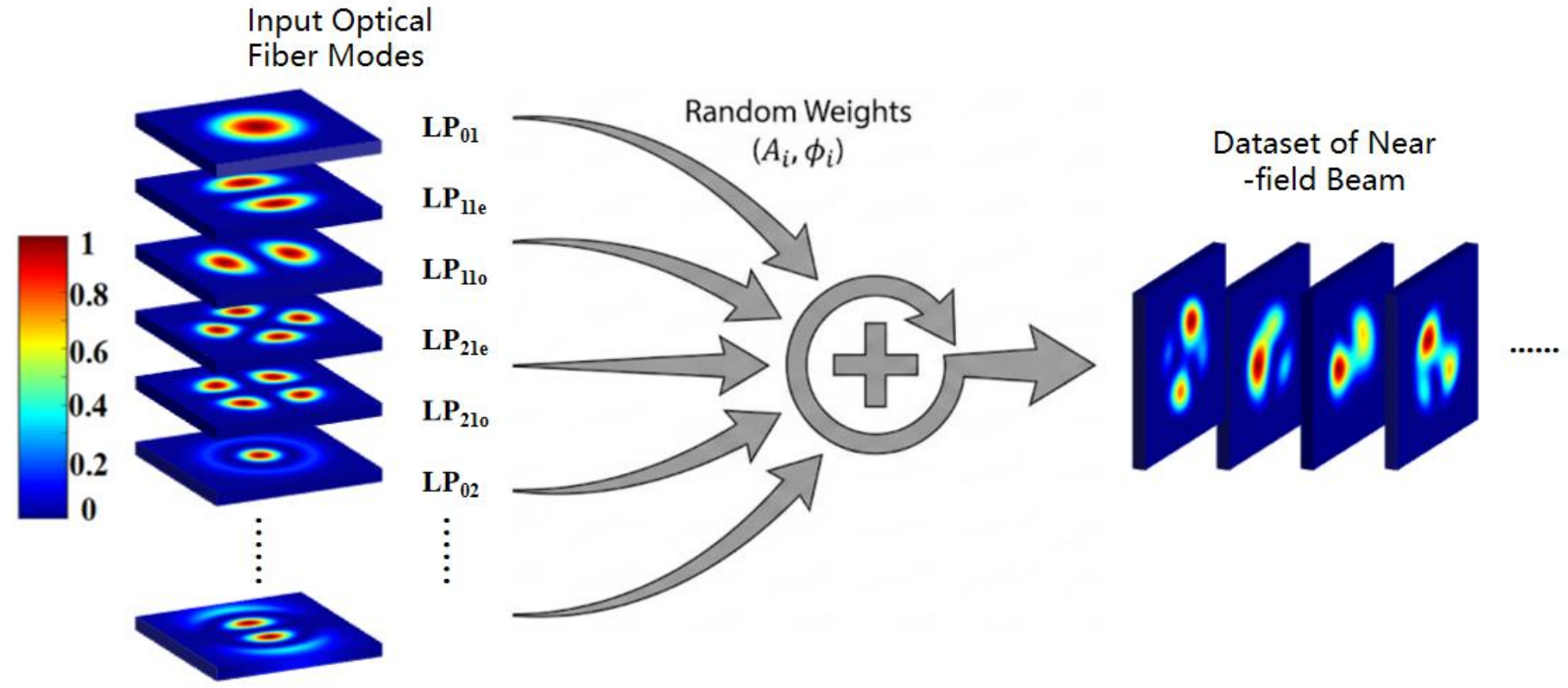

Fig. 3. Schematic representation of the sample generation process. The near-field beam profiles are numerically simulated by the coherent superposition of the LP modes with random amplitudes and phases.

As shown in Fig. 3, we simulated the output of a fiber laser as a coherent superposition of multiple linearly polarized (LP) modes. The complex optical field $E\ (x,\ y)$ at the fiber output facet can be expressed as a linear combination of the normalized orthogonal mode basis functions:

$$E(x,y) = \sum_{i=1}^{N} c_i\, \psi_i(x,y) = \sum_{i=1}^{N} A_i\, \psi_i(x,y)\, exp(j\phi_i), \tag{1}$$

where $\psi_i(x,y)$ represents the spatial distribution of the $i$-th LP mode, and $N$ denotes the total number of modes considered. To emulate the dynamic output of a multimode fiber laser, we selected the first twenty distinct LP modes ($LP_{01}$, $LP_{11e}$, $LP_{11o}$, $LP_{21e}$, $LP_{21o}$, $LP_{02}$, $LP_{31e}$, $LP_{31o}$, $LP_{12e}$, $LP_{12o}$, $LP_{41e}$, $LP_{41o}$, $LP_{22e}$, $LP_{22o}$, $LP_{03}$, $LP_{51e}$, $LP_{51o}$, $LP_{32e}$, $LP_{32o}$ and $LP_{13e}$) as the basis set. The complex coefficient $c_i$ is determined by the amplitude $A_i$ and the relative phase $\phi_i$ of each mode.

To ensure the conservation of total energy, the mode weights are normalized such that the sum of their squared amplitudes equals unity:

$$\sum_{i=1}^{N} |c_i|^2 = \sum_{i=1}^{N} A_i^2 = 1. \tag{2}$$

During dataset generation, the amplitudes $A_i$ were randomly sampled and subsequently normalized according to Eq. (2), while the phases $\phi_i$ were uniformly distributed in the range of [0, $2\pi$]. This random superposition mimics the mode coupling and dynamic instability observed in real fibers.

To automatically label each generated beam profile with its corresponding beam quality factor, we utilized the rigorous second-order moment definition derived from the generalized beam propagation theory (Yoda's formulation) [46]. The beam quality is quantified along a generalized transverse coordinate $u$ (where $u$ represents either $x$ or $y$). According to Yoda's formulation, the $M^2$ factor is determined by the spatial variance $\sigma_u^2$ and two additional beam parameters $A_u$ and $B_u$, which are calculated from the complex optical field $E(x,y)$ as follows:

$$\sigma_u^2 = \iint (u-\bar{u})^2 |E(x,y)|^2 dxdy, \tag{3}$$

$$A_u = \iint \left(u-\bar{u}\right)\left[E(x,y)\frac{\partial E^*(x,y)}{\partial u} - E^*(x,y)\frac{\partial E(x,y)}{\partial u}\right] dxdy, \tag{4}$$

$$B_u = \iint \left|\frac{\partial E(x,y)}{\partial u}\right|^2 dxdy + \frac{1}{4}\left\{\iint \left[E(x,y)\frac{\partial E^*(x,y)}{\partial u} - E^*(x,y)\frac{\partial E(x,y)}{\partial u}\right] dxdy\right\}^2. \tag{5}$$

Where $\bar{u}$ is the centroid of the beam and the field $E(x,y)$ is assumed to be normalized. Consequently, the ground-truth $M^2$ factor associated with the generated beam profile is computed as:

$$M_u^2 = \sqrt{4B_u\sigma_u^2 + A_u^2}\,, \tag{6}$$

and the effective $M^2$ factor is calculated by

$$M^2 = \sqrt{M_x^2 M_y^2}. \tag{7}$$

By iteratively randomizing the mode weights and phases under the normalization constraint, we generated a comprehensive dataset of near-field beam images paired with precise $M^2$ labels, providing a robust foundation for training the neural network.

### 2.3. Data acquisition and preprocessing

The optical pulse duration generated by the MZM is $\tau_{\text{pulse}} = 0.5$ ns, and the pulse interval is set to $\tau_{\text{internal}} = 10$ ns, resulting in a measurement rate of 100 MHz. The data transmission process is illustrated in Fig. 4. Modulated optical pulses characterized by varying beam quality factors are coupled into the MMF. Due to intermodal interference and crosstalk during transmission, distinct speckles are formed at the output of the MMF. After these speckles are spatially sampled by the seven cores of the MCF, the parallel optical signals must be serialized to realize single-pixel detection. This spatial-to-temporal conversion is achieved through a precisely designed fiber delay line array connected to the fan-out device. As depicted in the system schematic, the seven spatial channels are coupled into independent single-mode fibers (SMFs) of varying lengths. This structure effectively maps the spatial position of each core to a specific time slot in the output signal.

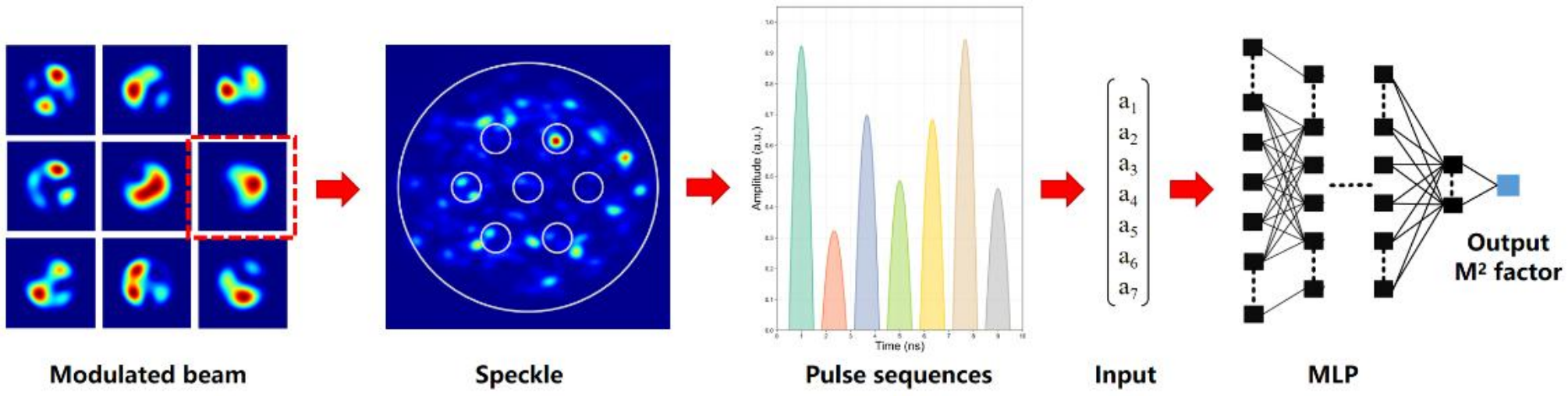

Fig. 4. Schematic of the data acquisition and processing framework.

To ensure that the optical pulses from different MCF cores arrive at the photodetector sequentially without temporal overlap, the fiber length $L_k$ for the $k$-th channel ($k$=1, 2, …, 7) is designed to follow an arithmetic progression:

$$L_k = L_1 + (k-1)\Delta L, \tag{8}$$

where $L_1$ is the reference length of the first channel, and $\Delta L$ is the incremental fiber length between adjacent channels. This length difference introduces a deterministic relative time delay $\Delta\tau$ between consecutive pulses:

$$\Delta\tau = \frac{n_{eff}\Delta L}{c}, \tag{9}$$

where $n_{eff}$ denotes the effective refractive index of the SMF core, and $c$ is the speed of light in a vacuum. To prevent inter-symbol interference (ISI), the selection of $\Delta L$ must strictly control the induced time delay $\Delta\tau$ to exceed the full duration $\tau_{\text{pulse}}$ of the optical pulse (including dispersion-induced broadening). Mathematically, this non-overlapping condition is expressed as $\Delta\tau > \tau_{\text{pulse}}$. Simultaneously, to ensure that the pulse train fits within one measurement cycle, the propagation delay difference between the shortest and longest fibers must satisfy the condition $7\,\Delta\tau < \tau_{\text{internal}}$. In our experiment, we selected a length increment of $\Delta L = 0.2$ m, resulting in $\Delta\tau = 1$ ns.

Upon detection by the high-speed PD, the serialized optical packet is converted into an analog electrical signal and digitized by the DSO. Data acquisition is synchronized with the pulse generation module via a common trigger signal from the AWG. For each laser pulse cycle, the DSO captures a waveform containing a sequence of seven distinct intensity peaks, which correspond one-to-one to the local light intensities at the seven cores of the MCF.

In the preprocessing stage, we implement a peak extraction algorithm to compress the raw waveform data. The peak voltages of the seven pulses are extracted to form a 7-dimensional feature vector x = $[a_1, a_2, \ldots, a_7]^T$. This vector serves as a compact digital fingerprint of the original modulated beam pattern. Finally, the extracted feature vector is normalized to eliminate the influence of total power fluctuations, serving as the standardized input for the subsequent neural network model. Through this pipeline, high-dimensional spatial beam information is efficiently compressed and serialized, thereby achieving ultrafast measurement throughput.

## 3. Results and discussion

### *3.1. Neural network architecture and training*

To reconstruct the beam quality factor from the compressed time-domain signals, a deep Multilayer Perceptron (MLP) was implemented as the regression model. The architecture consists of an input layer, eleven hidden layers, and an output layer. The input layer comprises 7 neurons, accepting the 7-dimensional feature vector x extracted from the serialized MCF pulses. To capture the highly non-linear mapping between the speckle-induced temporal features and the beam quality under noisy conditions, the network depth was increased to 11 hidden layers, with each layer utilizing the Rectified Linear Unit (ReLU) activation function. The output layer contains a single neuron with a linear activation function to predict the continuous $M^2$ value.

Data collection was performed experimentally, acquiring a total of 1,000,000 samples covering an $M^2$ factor range from 1.0 to 5.0. The dataset was randomly partitioned into a training set, a validation set, and a testing set in a ratio of 8:1:1 (800,000 for training, 100,000 for validation, and 100,000 for testing). Crucially, due to the stochastic influence of system noise and environmental perturbations, such as slight fiber vibrations or temperature fluctuations during the acquisition process, the temporal pulse sequences captured for the same modulated beam image exhibit inherent variations. This diversity in the dataset is instrumental, as it allows the neural network to learn robust feature representations that are invariant to environmental factors, thereby enhancing the model's generalization capability in real-world scenarios.

The network was implemented using the PyTorch framework. We employed the Mean Squared Error (MSE) as the loss function for optimization, while the prediction performance was quantified using the Mean Relative Error (MRE), defined as:

$$MRE = \frac{1}{N}\sum_{i=1}^{N} \left| \frac{y_i - \hat{y}_i}{y_i} \right| \times 100\%. \tag{10}$$

where $N$ is the number of test samples, $y_i$ is the ground truth, and $\hat{y}_i$ is the predicted value. The weights were updated using the Adam optimizer with an initial learning rate of 0.001.

### *3.2. Performance evaluation and discussion*

The performance of the FLASH and the deep regression model was rigorously evaluated. Figure.5 illustrates the evolution of the percentage error curves for both the training and validation sets over 1000 training epochs. The learning dynamics exhibit a distinct two-phase behavior: initially, the error curves decay precipitously, indicating that the network is efficiently extracting dominant feature representations from the serialized time-domain speckles. Subsequently, the curves transition into a saturation phase, converging to a stable minimum after approximately 500 epochs. Crucially, the validation error curve maintains a close alignment with the training error curve throughout the entire optimization process. This tight coupling suggests that the 11-layer MLP possesses strong generalization capabilities, effectively mitigating the risk of overfitting despite the high dimensionality of the feature space and the complexity of the non-linear mapping.

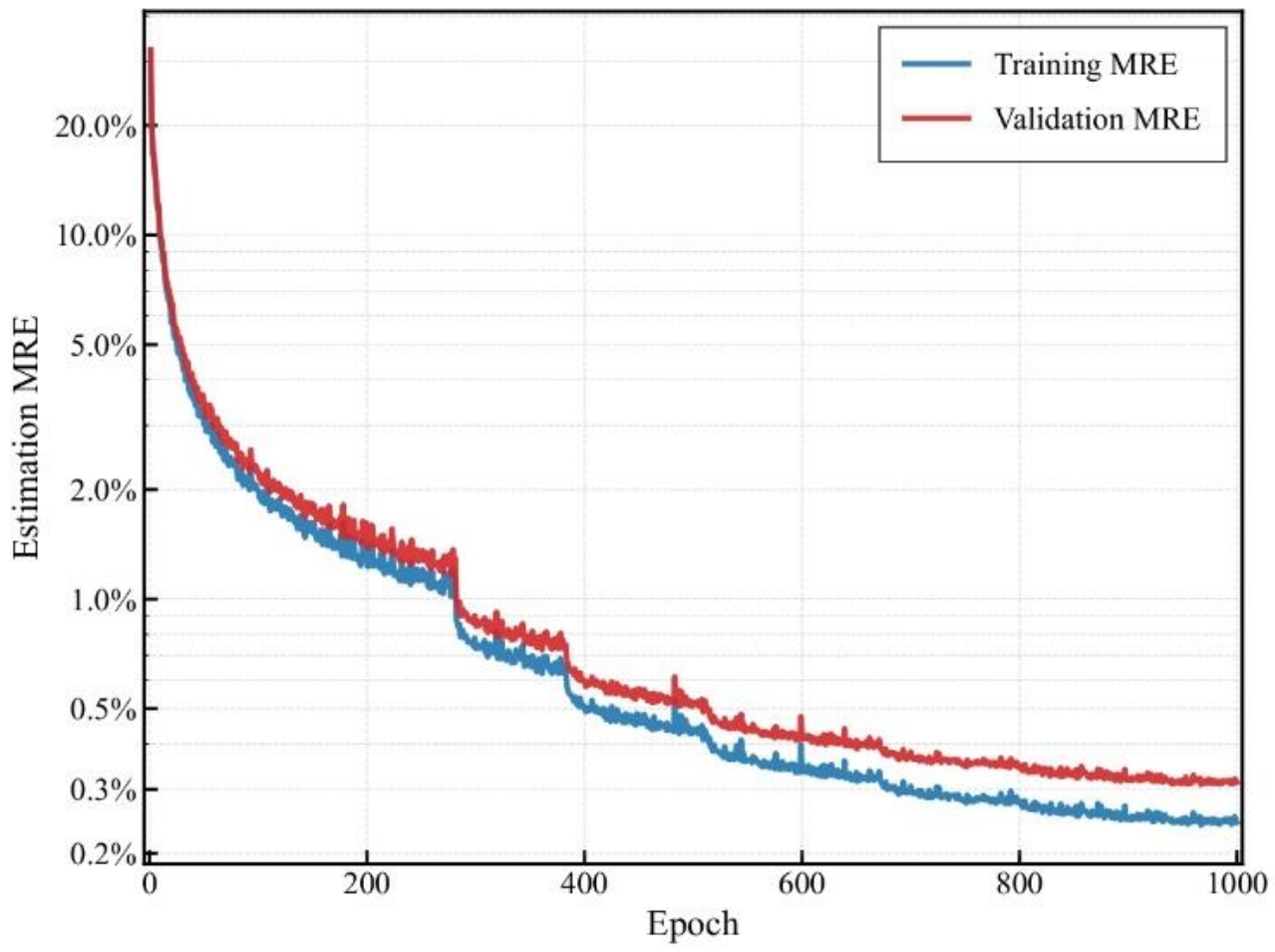

Fig. 5. Training and validation percentage error curves.

To further assess the prediction fidelity and quantify the system's uncertainty, the trained model was deployed on the independent testing set comprising 100,000 samples. Figure.6 displays the prediction performance and uncertainty analysis. The scatter plot reveals a remarkable degree of linearity between the predicted $M^2$ factors and the ground-truth values across the entire range from 1.0 to 5.0. The data points form a compact distribution along the ideal y = x diagonal, with a Pearson correlation coefficient exceeding 0.99.

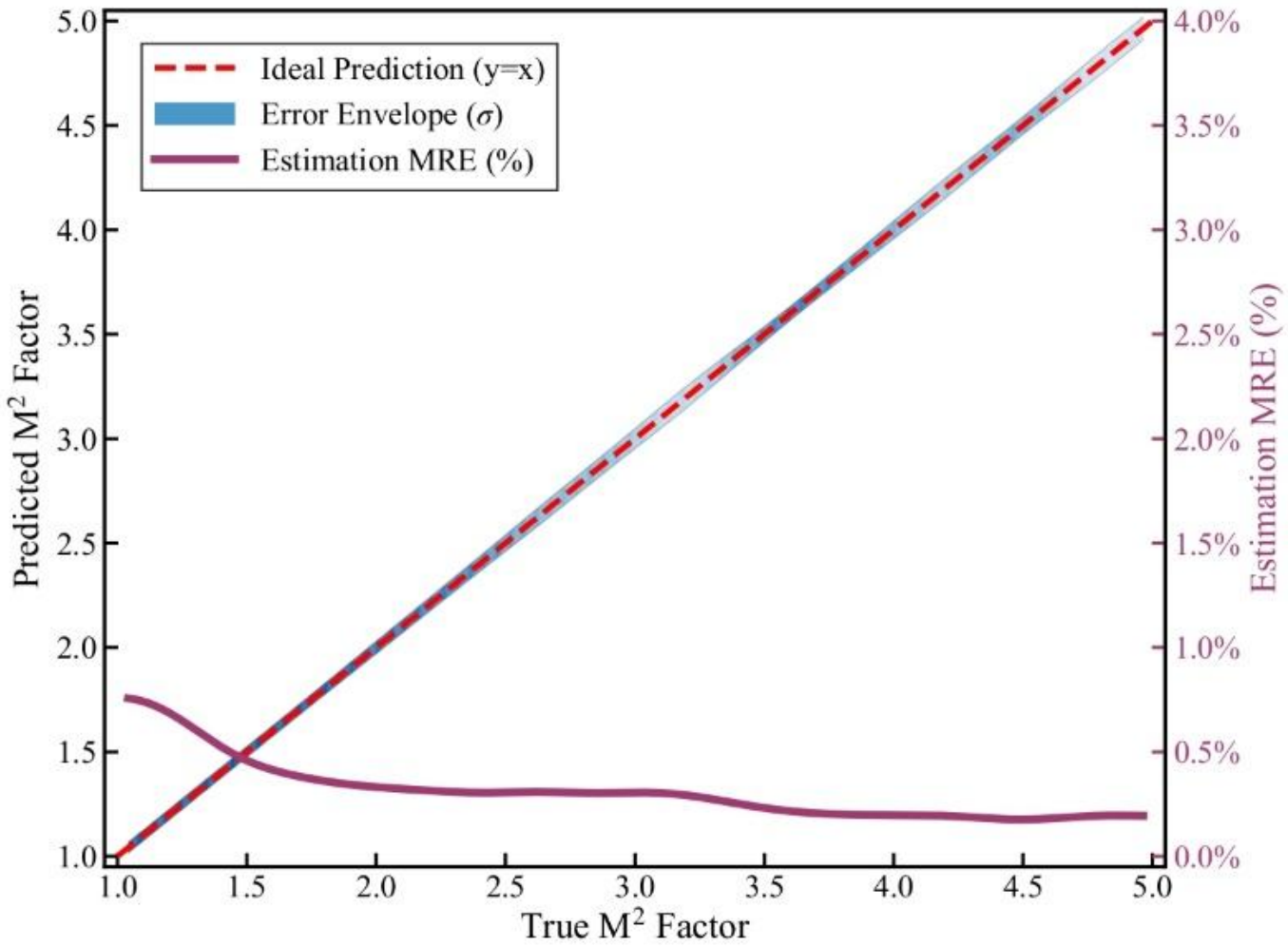

Fig. 6. Prediction performance and uncertainty analysis.

Of particular significance is the blue error envelope visible in Fig. 6, which represents the variance of the predicted values for a given ground-truth label. Since our dataset incorporates inherent stochastic variations in the pulse sequences—arising from system noise and environmental perturbations during data acquisition—this envelope effectively visualizes the model's prediction uncertainty. The fact that this error envelope tightly surrounds the ideal prediction line signifies high precision and robustness; it implies that the neural network has successfully learned to decouple the intrinsic beam quality information from environmental noise.

Quantitatively, the relative estimation error fluctuates slightly across the measurement range but remains consistently at a low level, with the majority of absolute errors confined within $\pm 0.01$. The overall system achieves an average MRE of 0.32%. This accuracy significantly outperforms conventional camera-based or numerical reconstruction methods, which typically suffer from errors over 1%. Combined with the 100 MHz measurement rate enabled by the spatial-to-temporal mapping, these results validate the reliability and superior performance of the FLASH for real-time, high-precision laser characterization.

### *3.3. Measurement performance under different numbers of cores*

Replacing a high-resolution camera with a discrete MCF substantially increases the measurement speed but inevitably reduces the amount of spatial information acquired. In our system, the millions of pixels typically captured by a CCD/CMOS sensor are compressed into just seven intensity data points. To quantify the impact of this drastic information reduction on the $M^2$ factor estimation accuracy, we evaluated the system's performance using different subsets of sampling cores. We retrained and tested the MLP model with input vector dimensions ranging from 1 to 7, corresponding to the number of active cores used for sampling.

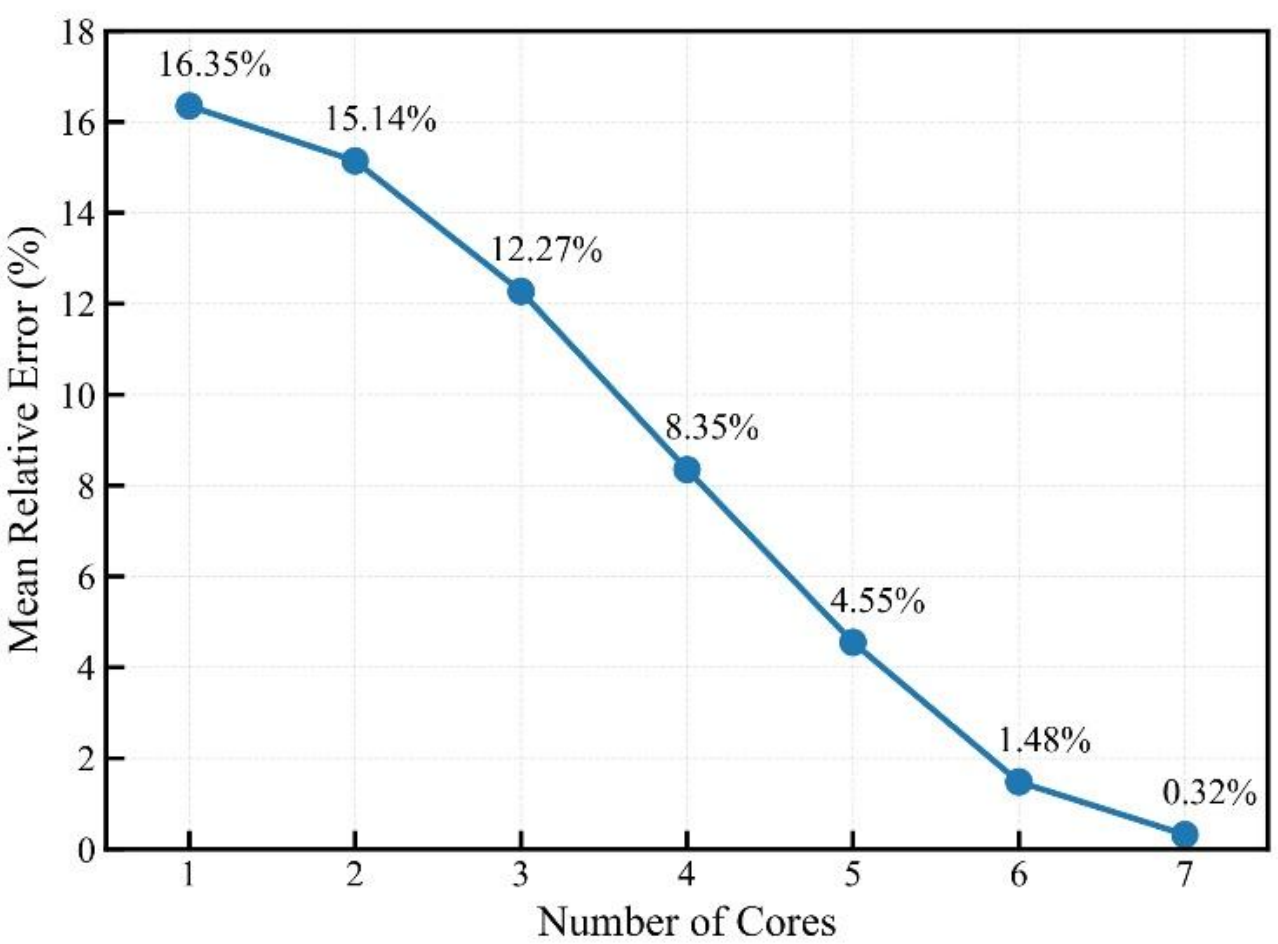

Fig. 7. Relationship between the $M^2$ estimation error (MRE) and the number of sampling cores.

The relationship between the estimation accuracy (represented by the MRE) and the number of sampling cores is illustrated in Fig. 7. As anticipated, the prediction error decreases monotonically as the number of sampling cores increases, indicating that acquiring more spatial information effectively resolves the ambiguity of the speckles. When only a single core is used, the MRE is relatively high, as one point is insufficient to capture the complex modal interference features. However, the accuracy improves significantly with the addition of cores. Notably, the improvement curve begins to saturate after the number of cores reaches six, suggesting that the marginal gain from additional sampling points diminishes. Considering the trade-off between measurement accuracy and system speed—since increasing the number of cores extends the total duration of the temporal pulse sequence and thus reduces the maximum repetition rate—we selected a seven-core MCF. This configuration achieves an optimal balance, providing high-precision $M^2$ estimation while maintaining an ultrafast measurement rate.

### 3.4. Measurement performance under varying modal complexities

In real-world high-power fiber lasers, the onset and progression of TMI involve chaotic energy transfer among a varying number of higher-order modes. To rigorously evaluate the robustness and the information-decoding capacity of our FLASH system, we systematically investigated its prediction performance under different modal complexities.

We constructed five distinct large-scale datasets by coherently superimposing the first 3, 5, 6, 8, and 10 linearly polarized modes. For each modal combination, 500,000 independent samples were generated and processed. The MLP model was independently retrained and tested for each dataset. Figure 8 illustrates MRE as a function of the number of mixed modes. Interestingly, the prediction error does not scale monotonically with the modal complexity but exhibits a distinct fluctuation. The recorded MRE values for 3, 5, 6, 8, and 10 modes are 0.33%, 0.53%, 0.32%, 0.16%, and 0.42%, respectively.

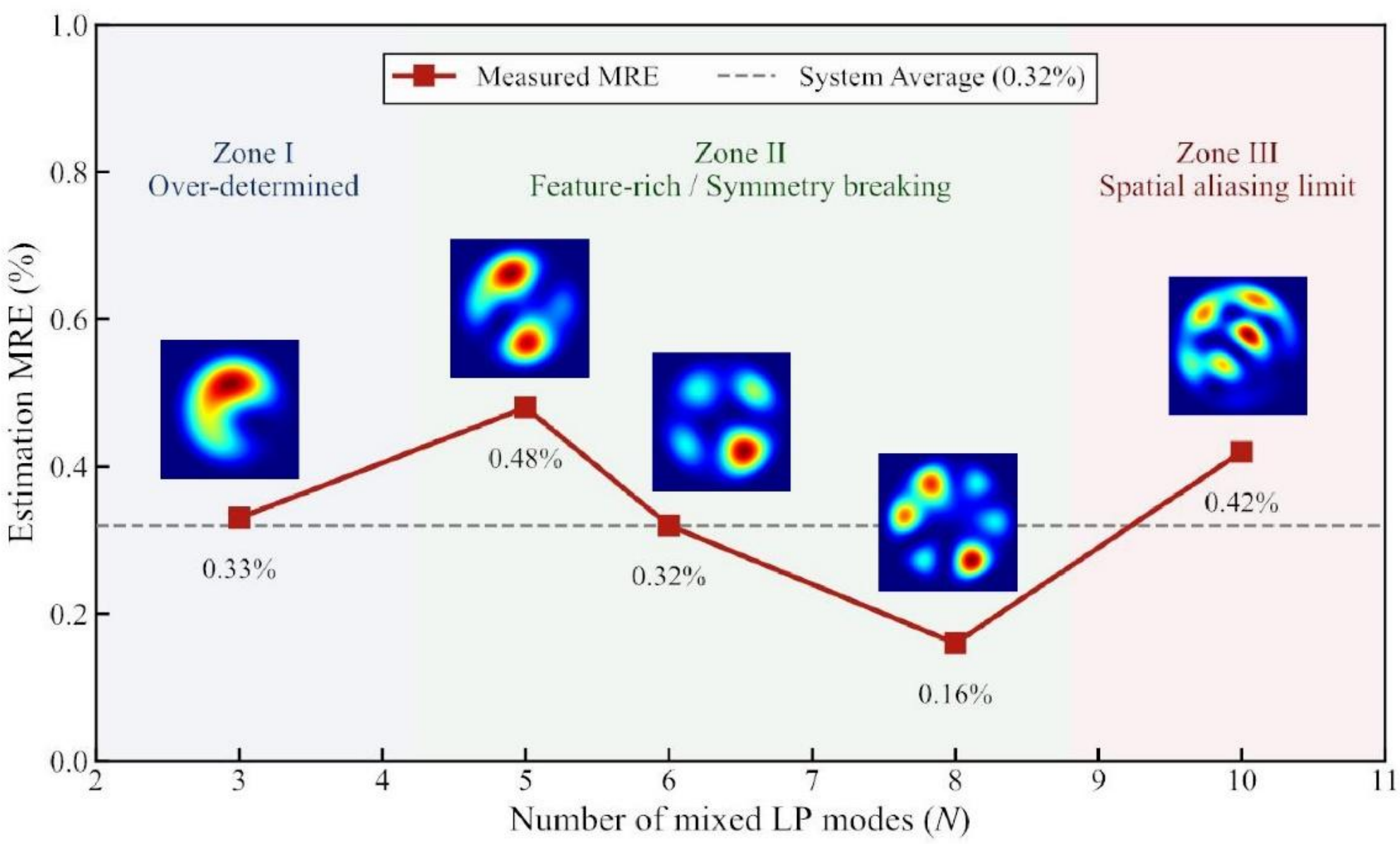

Fig. 8. Measurement performance under varying modal complexities.

This non-linear error dynamic reveals a fascinating physical interplay between spatial feature richness and sampling aliasing. For low mode counts (N=3), the limited degrees of freedom make the regression an easily determined problem. At N=5, the physical degrees of freedom slightly exceed the 7-core sampling capacity, introducing ambiguity that causes a local error peak. Surprisingly, adding more modes (N=6 and N=8) significantly reduces the error. The inclusion of specific higher-order modes introduces distinct and high-contrast interference cross-terms. These features break the spatial symmetries of lower-order superpositions and provide the neural network with richer, highly separable data representations, which actually enhances the regression accuracy.

However, as the mode count reaches 10, the speckle grains become extremely fine. The fixed 42-mm core pitch of the MCF leads to spatial aliasing where critical high-frequency spatial information falls into the blind spots between the sampling cores. This structural information loss ultimately causes the error to rise again. Despite this complex physical interplay, the maximum MRE remains exceptionally low at approximately 0.48%. This comprehensive evaluation proves that our system possesses sufficient channel capacity to unambiguously decode complex multimode interference and is highly robust for monitoring diverse dynamic states in fiber lasers.

## 4. Conclusion

In summary, we have proposed and experimentally validated FLASH, an ultrafast beam quality monitoring paradigm that effectively breaks the dimensional and temporal barriers of traditional optical measurements. By leveraging a multimode fiber for high-dimensional spatial encoding and a precisely designed multicore fiber delay line array for temporal serialization, we successfully mapped complex two-dimensional spatial beam profiles into one-dimensional high-speed temporal pulse sequences. Coupled with a deep multi-layer perceptron to decipher the complex intermodal interference signatures, the system robustly translates sparse serialized intensity data into highly accurate beam quality predictions.

Our experimental results demonstrate an unprecedented measurement rate of 100 MHz with an exceptionally low mean relative error of 0.32% across an $M^2$ factor range from 1.0 to 5.0. This achievement represents a staggering five-order-of-magnitude improvement in measurement speed compared to standard camera-based profiling techniques without compromising diagnostic precision. More importantly, FLASH functions as a transformative spatial oscilloscope. It successfully bridges the long-standing gap between slow spatial imaging and the ultrafast demands of modern laser physics.

Looking forward, this technique transcends the conventional scope of steady-state beam monitoring to provide an indispensable diagnostic tool for capturing nanosecond-level transient spatial evolutions. This capability unlocks the black box of complex multimode nonlinear physics, enabling the direct observation of spatio-temporal mode-locking and transient beam self-cleaning. Furthermore, its ultra-high bandwidth and low latency lay a solid hardware foundation for intelligent adaptive control. It empowers real-time closed-loop feedback to suppress TMI in large-mode-area fiber lasers and compensate for plasma-induced aberrations in advanced precision manufacturing. Finally, this paradigm offers a core technical means to combat high-frequency atmospheric vortex perturbations in free-space transmission. Ultimately, the proposed spatial-to-temporal mapping paradigm provides a highly practical and scalable diagnostic framework for the next generation of high-power and multimode laser systems.

**Acknowledgements.** This study was supported by the National Natural Science Foundation of China (62225110, 61931010), Major Program (JD) of Hubei Province (2023BAA013), and Hubei Provincial Natural Science Foundation of China (2025AFB008).

**Author contributions.** Junjie Qiu, Hao Wu and Ming Tang conceived and designed the work. Junjie Qiu, Yuxuan Xiong and Xuchen Hua acquired the data and performed the experiments. Junjie Qiu played an important role in interpreting the results. Junjie Qiu drafted the manuscript, Hao Wu and Ming Tang revised the manuscript. All authors approved the final version of the manuscript.

**Data Availability.** Data underlying the results presented in this paper are not publicly available at this time but may be obtained from the authors upon reasonable request.

**Conflict of interests.** The authors declare no conflicts of interest.